\documentstyle[11pt,epsf]{article}
\newcommand{\half}{\mbox{\small{$\frac{1}{2}$}}} 
\newcommand{\Nf}{N_{\!f}} 
\newcommand{\Nff}{\tilde{N}_{\!f}} 
\begin{document}
\title{Anomalous dimension of gluonic operator in polarized deep inelastic
scattering at $O(1/N_{\! f})$}  
\author{J.F. Bennett \& J.A. Gracey, \\ Theoretical Physics Division, \\ 
Department of Mathematical Sciences, \\ University of Liverpool, \\ 
Peach Street, \\ Liverpool, \\ L69 7ZF, \\ United Kingdom.}
\date{} 
\maketitle 
\vspace{5cm}
\noindent
{\bf Abstract.} We compute the $O(1/\Nf)$ correction to the predominantly 
gluonic flavour singlet twist-$2$ anomalous dimension used in polarized deep 
inelastic scattering. It is consistent with known two loop perturbation theory 
and we determine the three loop contribution at $O(1/\Nf)$. The treatment of 
the $\epsilon$-tensor in the large $\Nf$ $d$-dimensional critical point 
formalism is also discussed.  

\vspace{-17cm} 
\hspace{13.5cm} 
{\bf LTH-424}  

\newpage 
There has been intense experimental activity in recent years into understanding
the origin of the spin of the nucleons. This is primarily because of the EMC 
effect, \cite{1}, which appeared to suggest that the proton spin was due to a 
sea quark effect with no significant contribution from the valence quarks. 
Consequently there has been theoretical interest in the problem. For example, 
one avenue of research has been to improve theoretical predictions from 
perturbative QCD by going to higher loop orders, \cite{2}. A problem related to
this concerns the construction of the perturbative sector of the operator 
product expansion used in polarized deep inelastic scattering which is 
important, for example, for current and future HERA experiments. Unlike the 
unpolarized sector, polarized calculations of twist-$2$ operator dimensions 
have only recently become available at two loops, \cite{3,4}, as a function of 
the operator moment, $n$, or equivalently as a function of $x$, the conjugate 
variable of the Mellin transform in the DGLAP formalism of Dokshitzer, Gribov, 
Lipatov, Altarelli and Parisi, \cite{5,6}. Although the two loop results of 
\cite{3,4} represent progress for polarized scattering, building on the one 
loop results of \cite{7}, there is clearly a need to push calculations as far 
as is possible. Indeed for unpolarized scattering the relevant twist-$2$ 
operator dimensions are available exactly at three loops for the even moments, 
$n$ $\leq$ $8$, \cite{8} (and additionally $n$ $=$ $10$ for the flavour 
non-singlet case, \cite{9}). Although the explicit $n$-dependence of the three 
loop result is the main goal, since it is needed to carry out the full two loop
renormalization group evolution, some progress has been made in that direction 
already. Recently, the finite parts of the two loop graphs have been determined
in \cite{10} as a function of $x$. These represent the first stage of a full 
three loop calculation since they will contribute to the final value when 
multiplied by one loop counterterms. Further, some information is available
from the large $\Nf$ expansion where the $1/\Nf$ leading order non-singlet
anomalous dimensions have been deduced to all orders in the coupling, 
explicitly as a function of $n$, \cite{11}, with $\Nf$ the number of quark 
flavours. (The method used is based on a critical point argument where one 
considers the theory in the neighbourhood of a $d$-dimensional non-trivial zero
of the $\beta$-function and computes critical exponents. The initial 
development was for the $O(N)$ $\sigma$ model in \cite{12}. Through the 
critical renormalization group equation these exponents are related to the 
perturbative renormalization group functions.) The restriction of \cite{11} to 
compare with the exact three loop non-singlet results of \cite{9} gave 
agreement where there is overlap. In anticipation of a full polarized 
calculation at three loops the large $\Nf$ unpolarized calculations have been 
extended in \cite{13} to the polarized case. However, in the calculation of 
\cite{13} only the dimension of the twist-$2$ flavour singlet operator which 
was predominantly fermionic in nature, was considered. The aim of this article 
is to complete the gap in the twist-$2$ singlet polarized sector by computing 
the first non-trivial large $\Nf$ contribution to the flavour singlet operator 
which is gluonic in character. In providing this result we will not only be 
able to check the recent two loop results but also be able to provide the {\em 
first} insight into the $n$-dependence of the three loop dimension, albeit at 
$O(1/\Nf)$. As the calculation of the unpolarized singlet gluonic operators was
treated at length in \cite{14}, we will be brief concerning the technical 
details of the integration and focus on the salient differences which arise in 
the polarized case.   

We recall that the basic twist-$2$ flavour singlet polarized operators are, 
\cite{7}, 
\begin{eqnarray}
{\cal O}^{\mu_1 \ldots \mu_n}_q &=& i^{n-1} {\cal S} \bar{\psi} \gamma^5 
\gamma^{\mu_1} D^{\mu_2} \ldots D^{\mu_n} \psi ~-~ \mbox{trace terms} 
\nonumber \\ 
{\cal O}^{\mu_1 \ldots \mu_n}_g &=& \half i^{n-2} {\cal S} 
\epsilon^{\mu_1\alpha\beta\gamma} \, \mbox{tr} \, G^a_{\beta\gamma} D^{\mu_2} 
\ldots D^{\mu_{n-1}} G^{a \,\mu_n}_{~~~~\,\alpha} ~-~ \mbox{trace terms}
\label{ops} 
\end{eqnarray} 
where $\psi$ is the quark field, $G^a_{\mu\nu}$ $=$ $\partial_\mu A^a_\nu$ $-$
$\partial_\nu A^a_\mu$ $+$ $f^{abc} A^b_\mu A^c_\nu$, $A^a_\mu$ is the gluon
field, $f^{abc}$ are the colour group structure functions, ${\cal S}$ 
denotes symmetrization over the Lorentz indices and 
$\epsilon_{\mu\nu\sigma\rho}$ is a totally antisymmetric four dimensional 
pseudotensor. As has been observed in \cite{7}, since ${\cal O}_q$ and 
${\cal O}_g$ have the same dimension, they will mix under renormalization in 
perturbation theory. This leads to a mixing matrix of renormalization constants
and hence anomalous dimensions which we denote by  
\begin{equation} 
\gamma_{ij}(a) ~=~ \left( 
\begin{array}{ll} 
\gamma_{qq}(a) & \gamma_{gq}(a) \\ 
\gamma_{qg}(a) & \gamma_{gg}(a) \\ 
\end{array} 
\right) 
\label{mixmat} 
\end{equation} 
where we use the same definition for the coupling constant, $a$, as \cite{8}. 
To three loops the explicit $\Nf$ dependence of each of the entries will be 
\begin{eqnarray}  
\gamma_{qq}(a) &=& a_1a + (a_{21}\Nff + a_{22})a^2 + (a_{31}\Nff^2 + a_{32}\Nff 
+ a_{33})a^3 + O(a^4) \nonumber \\ 
\gamma_{gq}(a) &=& b_1a + (b_{21}\Nff + b_{22})a^2 + (b_{31}\Nff^2 + b_{32}\Nff 
+ b_{33})a^3 + O(a^4) \nonumber \\ 
\gamma_{qg}(a) &=& c_1 \Nff a + c_2\Nff a^2 + (c_{31}\Nff^2 + c_{32}\Nff 
+ c_{33})a^3 + O(a^4) \nonumber \\ 
\gamma_{gg}(a) &=& (d_{11}\Nff + d_{12})a + (d_{21}\Nff + d_{22})a^2 
+ (d_{31}\Nff^2 + d_{32}\Nff + d_{33})a^3 + O(a^4)  
\end{eqnarray} 
where $\Nff$ $=$ $T(R)\Nf$ and the coefficients $a_{ij}$, $b_{ij}$, $c_{ij}$ 
and $d_{ij}$ will depend on $n$ and the colour group Casimirs. This form will 
be important when we apply the critical point formalism to deduce the 
$O(1/\Nf)$ anomalous dimensions. Moreover, for concreteness and to clarify the 
conventions we use, we recall the one loop polarized results of Ahmed and Ross 
are, \cite{7}, 
\begin{eqnarray} 
a_1 &=& C_2(R) \left[ 4 S_1(n) ~-~ 3 ~-~ \frac{2}{n(n+1)} \right] ~~~,~~~  
b_1 ~=~ -~ \frac{2(n+2)C_2(R)}{n(n+1)} \nonumber \\
c_1 &=& -~ \frac{4(n-1)T(R)}{n(n+1)} ~~~,~~~  
d_{11} ~=~ \frac{4}{3} T(R) 
\end{eqnarray} 
where the finite sums $S_l(n)$ are defined by $S_l(n)$ $=$ $\sum_{r=1}^n 
1/r^l$. Although we have indicated the form of the mixing matrix in 
(\ref{mixmat}), one could in principle consider a basis where it is diagonal. 
Indeed it turns out that in the $1/\Nf$ formalism of \cite{13} that that method
produces the anomalous dimensions of the twist-$2$ eigenoperators of 
(\ref{mixmat}). This is primarily due to the fact that the canonical dimensions
of the operators of QCD in perturbation theory are the same, whereas computing 
at the fixed point of QCD in $d$-dimensional spacetime, where the large $\Nf$ 
expansion is constructed, the canonical dimensions are different. Indeed it is 
elementary to deduce this difference is $O(\epsilon)$ where $d$ $=$ $4$ $-$ 
$2\epsilon$. Further, it was shown in \cite{13} that the critical exponent of 
the operator ${\cal O}_f$ corresponded to the eigen-anomalous dimension of 
(\ref{ops}) with a predominantly fermionic content evaluated at the non-trivial
zero of the $\beta$-function. Likewise in \cite{14}, it was demonstrated that 
the remaining operator corresponded to the predominantly gluonic operator. More
concretely, diagonalizing (\ref{mixmat}) produces the eigen-anomalous 
dimensions  
\begin{equation} 
\lambda_\pm(a) ~=~ \frac{1}{2} ( \gamma_{qq} + \gamma_{gg} ) ~\pm~ 
\frac{1}{2} \left[ ( \gamma_{qq} - \gamma_{gg} )^2 + 4 \gamma_{qg}\gamma_{gq} 
\right]^{\half} 
\label{eigdim} 
\end{equation}  
The critical value of the QCD coupling has been given before in \cite{14} and 
is 
\begin{eqnarray}
a_c &=& \frac{3\epsilon}{4T(R)\Nf} ~+~ \left[ \frac{33}{16}C_2(G) \epsilon 
{} ~-~ \left( \frac{27}{16}C_2(R) + \frac{45}{16}C_2(G) \right) \epsilon^2 
\right. \nonumber \\ 
&& \left. +~ \left( \frac{99}{64}C_2(R) + \frac{237}{128}C_2(G) \right) 
\epsilon^3 ~+~ \left( \frac{77}{64}C_2(R) + \frac{53}{128}C_2(G) \right) 
\epsilon^4 \right] \frac{1}{T^2(R)\Nf^2} 
\label{critg}
\end{eqnarray}
where we have included the contribution from the recently computed four loop
$\beta$-function, \cite{15}. Therefore substituting (\ref{critg}) in 
(\ref{eigdim}) and expanding in powers of $1/\Nf$, we find that the gluonic 
dimension to third order in $\epsilon$ is given by  
\begin{eqnarray} 
\lambda_+(a_c) &=& \frac{3 d_{11} \epsilon}{4} ~+~ \frac{1}{T(R)\Nf} \left[ 
\frac{3}{4} \left( d_{12} + \frac{b_1c_1}{d_{11}} + \frac{11}{4}C_2(G) d_{11} 
\right) \epsilon \right. \nonumber \\ 
&&+~ \left. \frac{9}{16} \left( d_{21} + \frac{b_{21}c_1}{d_{11}}
- \left( 3C_2(R) + 5C_2(G) \right) d_{11} \right) \epsilon^2 \right. 
\nonumber \\ 
&&+~ \left. \frac{27}{64} \left( d_{31} + \frac{b_{31}c_1}{d_{11}}
+ \left( \frac{11}{3}C_2(R) + \frac{79}{18} C_2(G) \right) d_{11} \right) 
\epsilon^3 \right] ~+~ O \left( \frac{\epsilon^4}{\Nf^2} \right) 
\label{criteigdim} 
\end{eqnarray} 

To reproduce this at $O(\epsilon^2)$ together with the subsequent terms at 
$O(1/\Nf)$ one applies the critical point formalism to the operator 
${\cal O}_g$ which exploits the fixed point equivalence of QCD with the 
non-abelian Thirring model, \cite{16}, at leading order in $1/\Nf$. As the 
method has been well documented, \cite{14}, we will comment only on those 
technical details of the computation which are peculiar to ${\cal O}_g$. Indeed
as the calculation virtually parallels that of the unpolarized case we refer 
the interested reader to \cite{14}. There details of the integration procedure 
for the Feynman diagrams have been recorded, including the construction and 
solution of the recurrence relations for deducing the contributions from 
subintegrals in the graphs. These have been implemented in the symbolic 
manipulation language {\sc form}, \cite{17}, which was used to perform the 
tedious algebra in the integration. The four contributing Feynman graphs for 
the $O(1/\Nf)$ part of (\ref{criteigdim}) are illustrated in fig. 1, where the 
operator insertion is indicated on an internal gluon line. We have used the 
same Feynman rule for the insertion and conventions as \cite{3}. 

One technical issue deserves detailed comment. As has been noted elsewhere, 
\cite{18}, one has to deal with the problem of treating an object, such as 
$\epsilon_{\mu\nu\sigma\rho}$, which is defined only in four dimensions, in a 
formalism which is manifestly $d$-dimensional. Such a situation arises not only
in large $\Nf$ calculations but also in explicit perturbative multiloop 
computations. Its resolution in the latter case is to treat products of two 
$\epsilon$-tensors, where they arise, as the determinant of a $4$ $\times$ $4$ 
matrix of $\eta_{\mu\nu}$ tensors, \cite{19}. As the latter can have a 
$d$-dimensional interpretation one can perform a calculation without reference 
to the original $\epsilon$-tensors. To compensate for the obvious discrepancy 
and to restore the Ward identity, one introduces an additional (finite)
renormalization constant, \cite{20}. This approach was developed for the 
treatment of the renormalization of the singlet axial vector current which has 
an anomaly, \cite{20,21}. A similar approach is adopted for $1/\Nf$ 
calculations, \cite{22}, which has already been applied to the calculation of 
the exponent corresponding to the anomalous dimension of the axial vector 
current, \cite{13}. For each graph of the present calculation, we project out 
the contribution to the exponent by multiplying by 
$\epsilon_{\mu\nu\sigma\rho}\Delta^\sigma p^\rho$, where $\Delta^\mu$ is a null
$4$-vector used to define the Feynman rules on the light cone and $p^\mu$ is 
the external momentum flowing through each diagram. Therefore each integral 
contains the product of two $\epsilon$-tensors and like \cite{20} we replace 
these by a determinant of $\eta_{\mu\nu}$ tensors and compute in 
$d$-dimensions. However, we also need to introduce a compensating factor since 
not only is $\epsilon_{\mu\nu\sigma\rho}$ four dimensional but the vectors 
$\Delta^\mu$ and $p^\mu$ are strictly physical objects and cannot be extended 
to $d$-dimensions, \cite{18}. In the split $\gamma$-algebra notation of 
\cite{18}, $\Delta^\mu$ $\equiv$ $\Delta^{\bar{\mu}}$ and $p^\mu$ $\equiv$ 
$p^{\bar{\mu}}$ where the bar denotes the four dimensional subspace of the 
$d$-dimensional extended spacetime, which corresponds to the physical 
spacetime in the limit as $d$~$\rightarrow$ $4$. To summarize we have replaced 
the correct $\epsilon$-tensor combination given on the right side of  
\begin{equation} 
\epsilon_{\mu\nu\sigma\rho} \epsilon_{\alpha\beta\gamma\delta} \Delta^\sigma
p^\rho \Delta^\gamma p^\delta ~=~ f(\mu) 
\epsilon_{\mu\nu\bar{\sigma}\bar{\rho}} 
\epsilon_{\alpha\beta\bar{\gamma}\bar{\delta}} \Delta^{\bar{\sigma}} 
p^{\bar{\rho}} \Delta^{\bar{\gamma}} p^{\bar{\delta}} 
\label{proj} 
\end{equation} 
by the object on the left side which is formally $d$-dimensional and involves
the $\eta$-determinant. The discrepancy in this procedure is represented by a 
compensating function of the spacetime dimension. We have computed this 
proportionality function, $f(\mu)$, by considering contractions of the 
$d$-dimensional object and comparing it with similar contractions of the object
restricted to $4$-dimensional spacetime. We found 
\begin{equation} 
f(\mu) ~=~ (2\mu-3)(\mu-1) 
\label{fval} 
\end{equation} 
To check that this is a correct procedure we computed the first graph of fig. 1
by using the determinant substitution approach and then compared the result 
with that obtained by direct calculation of the integral. Such a computation is
possible in this case since the graph is in effect a series of chain integrals 
in the language of \cite{12}. It turns out that by inspecting each of the other
graphs a combination which is effectively the same as (\ref{proj}) arises and 
therefore (\ref{fval}) is used for each of the graphs. Moreover, this argument 
was also used to justify the inclusion of a similar factor in the large $\Nf$
calculation of the anomalous dimension of the singlet axial vector current, 
\cite{13}. The final expression for that case was consistent with three loop 
perturbation theory, \cite{21,19}. 

In light of these remarks, we can now quote the final value we obtained for 
$\lambda_{+,1}(a_c)$ where $\lambda_+(a_c)$ $=$ $\sum_{i=0}^\infty 
\lambda_{+,i}(a_c)/\Nf^i$. After evaluating the contribution from each graph we
find 
\begin{eqnarray} 
\lambda_{+,1}(a_c) &=& -~ \frac{(n + 2)(n - 1)\Gamma(n+2-\mu)\Gamma(\mu + 1) 
C_2(R)\eta_1^{\mbox{o}}}{(\mu - 2)^2(n + 1)\Gamma(2 - \mu)\Gamma(\mu + n)n 
T(R)} \nonumber \\ && \nonumber \\ && +~ \frac{2\mu(\mu - 1) S_1(n) C_2(G) 
\eta_1^{\mbox{o}}} {(2\mu - 1)(\mu - 2) T(R)} \nonumber \\ && \nonumber \\  
&& -~ [4\mu^3n^2 + 4\mu^3n - 8\mu^3 - 8\mu^2n^2 - 8\mu^2n + 16\mu^2 + 5\mu n^2 
\nonumber \\ 
&& ~~~ + 5\mu n - 9\mu - n^2 - n + 2] \Gamma(n+2-\mu) \Gamma(\mu)\mu C_2(G)  
\eta_1^{\mbox{o}} \nonumber \\ 
&& ~~~~~ /[8(2\mu - 1)(\mu - 1)^3(n + 1) \Gamma(3 - \mu)\Gamma(\mu + n)n T(R)] 
\nonumber \\ && \nonumber \\  
&& -~ [32\mu^4 - 4\mu^3n^2 - 4\mu^3n - 120\mu^3 + 16\mu^2n^2 + 16 \mu^2n 
\nonumber \\ 
&& ~~~ + 160\mu^2 - 20\mu n^2 - 20\mu n  - 89\mu + 8n^2 + 8n + 18] \mu C_2(G) 
\eta_1^{\mbox{o}} \nonumber \\ 
&& ~~~~~ /[8(2\mu - 1)(\mu - 1)^3(\mu - 2)(n + 1)n T(R)] 
\label{res} 
\end{eqnarray} 
where we have used {\sc reduce}, \cite{23}, to handle tedious amounts of 
algebra. The quantity $\eta_1^{\mbox{o}}$ is defined by 
\begin{equation} 
\eta_1^{\mbox{o}} ~=~ \frac{(2\mu-1)(\mu-2)\Gamma(2\mu)} 
{4\Gamma^2(\mu)\Gamma(\mu+1)\Gamma(2-\mu)} 
\end{equation}  
There are various checks on the validity of (\ref{res}). First, we find that 
expanding (\ref{res}) in powers of $\epsilon$ it agrees with two loop
perturbation theory, \cite{7,3,4}. From one point of view this justifies the 
use of (\ref{proj}), since (\ref{fval}) is $\epsilon$-dependent and therefore 
will affect the form of the $O(\epsilon^2)$ terms of (\ref{res}) when it is 
expanded. Second, if one substitutes $n$ $=$ $1$ in (\ref{res}), we find that 
it vanishes identically. This is consistent with the observations of 
\cite{24,25}, where it was shown that the axial anomaly forces the $n$ $=$ $1$
gluonic operator to have vanishing anomalous dimension. Therefore we are 
confident that (\ref{res}) is correct and can make an observation on the form 
of the $3$-loop contribution to the gluonic eigen-anomalous dimension. By 
expanding (\ref{res}) to $O(\epsilon^3)$ and using (\ref{criteigdim}), we find 
that 
\begin{eqnarray} 
d_{31} ~+~ \frac{b_{31}c_1}{d_{11}} &=& -~  
\frac{64(7n^2 + 7n + 3)(n + 2)(n - 1) S_1(n) C_2(R)}{9(n + 1)^3n^3} 
\nonumber \\ 
&& +~ \frac{64(n + 2)(n - 1) S_1^2(n) C_2(R)}{3(n + 1)^2n^2} \nonumber \\ 
&& -~ 4[33n^8 + 132n^7 + 142n^6 - 36n^5 - 263n^4 - 312n^3 \nonumber \\ 
&& ~~~~ + 280n^2 + 408n + 144] C_2(R)/[27(n + 1)^4n^4] \nonumber \\ 
&& -~ \frac{8(8n^4 + 16n^3 - 19n^2 - 27n + 48) S_1(n) C_2(G)}{27(n + 1)^2n^2}  
\nonumber \\ 
&& -~ \frac{2(87n^6 + 261n^5 + 249n^4 + 63n^3 - 76n^2 - 64n - 96) C_2(G)} 
{27(n + 1)^3n^3}  
\label{3loop} 
\end{eqnarray} 
Evaluating this at $n$ $=$ $1$, one obtains a value which, from 
(\ref{criteigdim}), is correctly proportional to the three loop coefficient of 
$a_c$. As an aid to future explicit perturbative calculations similar to 
\cite{8}, we have evaluated (\ref{3loop}) for odd $n$, $n$ $\leq$ $23$, and 
listed the results in table 1. Unlike in the unpolarized calculations, here it 
is the odd moments which are important since the original operator does not 
exist when $n$ is even as can be seen, for example, from the Feynman rules, 
\cite{3}. A further remark on (\ref{3loop}) concerns the status of its $C_2(G)$
sector. In the unpolarized case, \cite{14}, it was noted that at $O(1/\Nf)$ 
$b_{31}$ will only involve contributions proportional to $C_2(R)$ and not 
$C_2(G)$. Assuming the situation is similar here, this implies that the terms 
involving $C_2(G)$ in (\ref{3loop}) are in fact the correct $n$-dependent part 
of the coefficient set $d_{31}$ of the original perturbative operator. Finally,
several general remarks on the overall structure are in order. First, the full 
expression is much simpler in form than that of the unpolarized case, 
\cite{14}. Indeed this was also observed in the fermionic singlet sector as 
well. Second, in the evaluation of each integral an extra $n$-dependent 
structure was present which was proportional to 
$\Gamma(\mu-1)\Gamma(n)/\Gamma(\mu+n)$.  This is similar to the unpolarized 
situation, \cite{14}, and like in that calculation, it also cancels in the 
final summation. 

To conclude with we note that all the $1/\Nf$ leading order terms are now
available for the anomalous dimensions of the twist-$2$ operators used in 
polarized deep inelastic scattering. The next stage in this $1/\Nf$ programme 
would be to consider the unpolarized flavour non-singlet sector and compute the
$O(1/\Nf^2)$ correction to the dimension of the twist-$2$ operator as a 
function of $n$.  

\vspace{1cm}  
{\bf Acknowledgements.} This work was carried out with the support of PPARC 
through a Postgraduate Studentship (JFB) and an Advanced Fellowship (JAG).

\newpage 
{\begin{table} 
\begin{center} 
\begin{tabular}{c||r|r} 
$n$ & $C_2(R)$ coefficient & $~~~~~~~C_2(G)$ coefficient \\ 
\hline  
& & \\ 
1 & $ - \, \frac{44}{9} $ & $ - \, \frac{158}{27} $ \\ 
& & \\ 
3 & $ - \, \frac{5527}{972} $ & $ - \, \frac{2342}{243} $ \\ 
& & \\ 
5 & $ - \, \frac{716824}{151875} $ & $ - \, \frac{341683}{30375} $ \\ 
& & \\ 
7 & $ - \, \frac{155752763}{34574400} $ & $ - \, \frac{4524703}{370440} $ \\ 
& & \\ 
9 & $ - \, \frac{2698556278}{602791875} $ & 
$ - \, \frac{29607251}{2296350} $ \\ 
& & \\ 
11 & $ - \, \frac{282328566167}{62758939320} $ & 
$ - \, \frac{1822332833}{135841860} $ \\ 
& & \\ 
13 & $ - \, \frac{5077270251071}{1120171987935} $ & 
$ - \, \frac{278760143099}{20142952830} $ \\ 
& & \\ 
15 & $ - \, \frac{915045986064491}{200354394240000} $ & 
$ - \, \frac{16575124393}{1167566400} $ \\ 
& & \\ 
17 & $ - \, \frac{189383384467690249}{41180834365746075} $ & 
$ - \, \frac{9360556077617}{645328543860} $ \\ 
& & \\ 
19 & $ - \, \frac{4489933300070894563}{970409113734060000} $ & 
$ - \, \frac{10746754033373}{727252911000} $ \\ 
& & \\ 
21 & $ - \, \frac{934060610086114519}{200816623282959684} $ & 
$ - \, \frac{3109748568599}{207034063236} $ \\ 
& & \\ 
23 & $~~ - \, \frac{128838538840521478537}{27573179435993423232} $ & 
$ - \, \frac{77700727191509}{5098296739536} $ \\ 
& & \\ 
\end{tabular} 
\end{center} 
\end{table} 
\begin{center} 
{Table 1. Coefficients of $\left[ d_{31} \right.$ $+$ $\left. 
b_{31}c_1/d_{11} \right]$ as a function of moment.} 
\end{center} } 

\newpage 
{\epsfysize=7cm 
\epsfbox{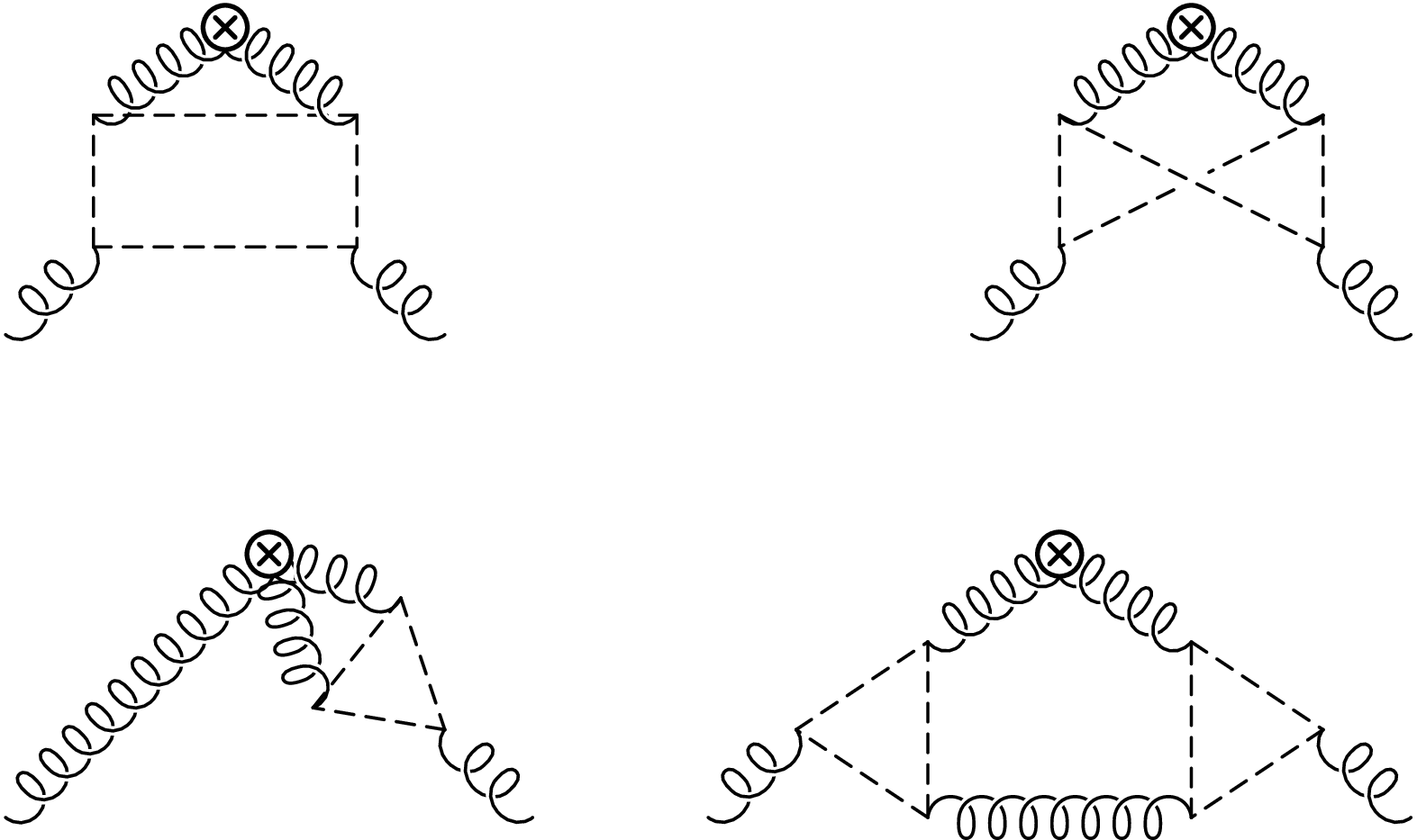}} 
\vspace{1cm} 
{\bf Fig. 1. Leading order diagrams for $\lambda_+(a_c)$.} 

\end{document}